\documentclass{aa}
\usepackage{txfonts}
\usepackage{epsf}
\begin{document}

\title{Shears from shapelets}
\author{Konrad Kuijken\inst{1}}
\institute{Leiden Observatory, PO Box 9513, 2300RA
  Leiden, The Netherlands}
\date{Received <date> / Accepted <date>}
\abstract
{
}
{
Accurate measurement of gravitational shear from images of distant
  galaxies is one of the most direct ways of studying the distribution
  of mass in the universe. We describe a new implementation of a
  technique for measuring shear that is based on the shapelets
  formalism.  } 
{
The shapelets technique describes PSF and observed images in
  terms of Gauss-Hermite expansions (Gaussians times polynomials).  It
  allows the various operations that a galaxy image undergoes before
  being registered in a camera (gravitational shear, PSF convolution,
  pixelation) to be modeled in a single formalism, so that intrinsic
  ellipticities can be derived in a single modeling step.  }
{
The resulting algorithm, and tests of it on idealized data as well as
  more realistic simulated images from the STEP project, are
  described. Results are very promising, with attained calibration
  accuracy better than four percent (1 percent for round PSFs) and PSF
  ellipticity correction better than a factor of 20. Residual
  calibration problems are discussed.  } 
{
}

\maketitle

\section{Introduction}

Weak gravitational lensing is recognized as a profitable way to study
the dark matter distribution in the universe, and with a series of
ever wider-field astronomical cameras coming online, very large weak
lensing surveys are being planned and performed. The inevitable source
of noise in weak lensing measurements is shape noise, caused by the
diversity of projected galaxy shapes on the sky. To beat down the
noise, very large numbers of galaxy ellipticities need to be averaged:
thus new lensing surveys such as the CFHT Legacy Survey (Hoekstra et
al.\ \cite{Hoekstra2005}), the CTIO survey (Jarvis et al.\
\cite{Jarvis2003}), the VIRMOS-Descart survey (van Waerbeke, Mellier
\& Hoekstra \cite{VanWaerbeke2005}) or the recently-approved
Kilo-Degree Survey (KIDS) on the VLT Survey Telescope, will contain
millions of background sources, which in principle should enable very
accurate shear measurements.

This averaging down of the shape noise through large number statistics
only makes sense if systematic errors can be controlled: the main one
is still the blurring of the source images by atmosphere, telescope
and detector pixels. The commonly-used Kaiser, Squires \&
Broadhurst~(\cite{Kaiser1995}; henceforth KSB) and Luppino \& Kaiser
(\cite{Luppino1997}) methods provide recipes for correcting these
effects, and are very successful. Nevertheless, they are based on an
idealized model of the effect of point spread function (PSF)
convolution on ellipticity, and it is possible to construct plausible
PSFs that it fails to correct properly (e.g., Hoekstra et
al.~\cite{Hoekstra1998}, Appendix~D). Therefore it seems unlikely that
the KSB recipes will deliver the factor of 10 to 100 improvement in
fidelity that will be required to exploit the new surveys (Erben et
al.\ \cite{Erben2001}).

A number of different approaches have been put forward to improve PSF
correction (Kuijken \cite{Kuijken1999}; Kaiser \cite{Kaiser2000};
Rhodes, Refregier \& Groth \cite{Rhodes2000}; Bernstein \& Jarvis \cite{Bernstein2002}; Refregier \& Bacon
\cite{RefregierBacon2003}; Mandelbaum et al.\
\cite{Mandelbaum2005}). In this paper we present a new
technique which combines elements from most of these.

The paper is organized as follows: in section 2 shear and ellipticity
are defined, and the effect of one on the other. Section 3 summarizes
the shapelets formalism, and describes how a shapelet description of a
source and its PSF can be used to generate an ellipticity estimate
that is useful for shear estimation. Section 4 substantiates the
approach with idealized tests of the algorithm. In section 5 a
software pipeline is presented that implements the full processing
chain from an astronomical image to a shear estimate. Results of
applying the pipeline to test data from the STEP project are shown in
Section 6. In Section 7 we compare with other techniques, and Section 8
gives the conclusions.

\section{Preliminaries}
\label{sec:prelim}
\subsection{Shear and Distortion}

Following the usual practice, we parameterize the effect of weak
gravitational lensing on a distant source in terms of a shear
$(\gamma_1,\gamma_2)$ and a convergence $\kappa$: the distorted image
$I_\mathrm{lensed}(x,y)$ is derived from the original $I(x,y)$ via the
transformation $I_\mathrm{lensed}(x,y) = I(x',y')$, where
\begin{eqnarray}
\nonumber
\left(\!\!\begin{array}{c}x'\\y'\end{array}\!\!\right) &=
\left(\begin{array}{cc} 1-\gamma_1-\kappa&-\gamma_2\\
-\gamma_2&1+\gamma_1-\kappa\end{array}\right)
\left(\!\!\begin{array}{c}x\\y\end{array}\!\!\right)
\\
&\equiv
(1-\kappa)
\left(\begin{array}{cc} 1-g_1&-g_2\\-g_2&1+g_1\end{array}\right)
\left(\!\!\begin{array}{c}x\\y\end{array}\!\!\right)
\label{eq:ellipdef}
\end{eqnarray}
The first matrix in this equation (the {\em distortion matrix})
represents the transformation from observed $(x,y)$ to undistorted
$(x',y')$ coordinates.

Without knowledge of the intrinsic source size, only the {\em
distortion} $(g_1,g_2)\equiv(\gamma_1,\gamma_2)/(1-\kappa)$, which
affects the shape of the source, can be measured (Schneider \& Seitz
\cite{Schneider1995}).

\subsection{Ellipticity}

We define the ellipticity of an object's image $I$ as follows:

Let $I_\mathrm{ell}$ be the model image with constant-ellipticity
isophotes that best approximates $I$. Then
the ellipticity $(e_1,e_2)$ of $I$ is defined such that a distortion
of $(-e_1,-e_2)$ makes $I_\mathrm{ell}$ circular.

The major axis position angle $\phi$ and the axis ratio $q$ of an
elliptical source are simply related to $(e_1,e_2)$:
\begin{equation}
q={1-e\over 1+e}\qquad\hbox{and}\qquad 
e_1=e\cos2\phi,\qquad e_2=e\sin2\phi
\end{equation}

This definition is similar to the one adopted by Bernstein \& Jarvis
(\cite{Bernstein2002}; henceforth BJ02), but does not explicitly force
a fit to an elliptical Gaussian.

As discussed by BJ02, expressing the shapes of objects in terms of
distortions $e$ has a practical advantage: in this formulation it is
simple to calculate the response of object shapes to small
distortions. An elliptical source with ellipticity $(e_1,e_2)$ that is
sheared by a small amount $(g_1,g_2)$ can be viewed as a circular
source that is sheared twice, first by $e_i$ and then by $g_i$, giving
a combined distortion matrix
\begin{eqnarray}
\nonumber
&
\left(\begin{array}{cc} 1-e_1&-e_2\\-e_2&1+e_1\end{array}\right)
\left(\begin{array}{cc} 1-g_1&-g_2\\-g_2&1+g_1\end{array}\right)
\qquad\qquad\qquad\qquad\qquad
\\
&=
\left(
\begin{array}{cc}
1-e_1-g_1+e_1g_1+e_2g_2 & -e_2-g_2+e_1g_2-e_2g_1\\
-e_2-g_2-e_1g_2+e_2g_1 & 1+e_1+g_1+e_1g_1+e_2g_2\\
\end{array}
\right).\quad
\end{eqnarray}
This matrix is no longer a pure distortion matrix (which would have to
be symmetric and of trace 2), but 
some algebra shows that this matrix can be decomposed into a
magnification, a rotation and a distortion:
\begin{eqnarray}
\nonumber
&
\left(\begin{array}{cc} 1-e_1&-e_2\\-e_2&1+e_1\end{array}\right)
\left(\begin{array}{cc} 1-g_1&-g_2\\-g_2&1+g_1\end{array}\right)
\qquad
\qquad
\qquad
\\
&=(1+K)
\left(\begin{array}{cc} 1&R\\-R&1\end{array}\right)
\left(\begin{array}{cc} 1-e_1-\delta_1&-e_2-\delta_2\\
-e_2-\delta_2&1+e_1+\delta_1\end{array}\right),
\end{eqnarray}
where, to first order in the distortion $g_i$, 
\begin{eqnarray}
K=&e_1g_1+e_2g_2\\
R=&e_1g_2-e_2g_1\\
\delta_1=&(1-e_1^2-e_2^2)g_1\\
\delta_2=&(1-e_1^2-e_2^2)g_2
\end{eqnarray}
Thus the action of a small distortion $g$ on an elliptical source with
ellipticity $e_i$ (according to the definition in
eq.~\ref{eq:ellipdef}) is equivalent to acting on a circular source
with, successively, a magnification, a rotation (neither of which
affect the shape of a circular source), and a distortion by
$e_i+\delta_i$.
 
Assuming now that we have an ensemble of elliptical sources, of random
orientations, so that before distortion $\langle e_i\rangle=0$ and $\langle
e_1^2\rangle+\langle e_2^2\rangle\equiv \langle e^2\rangle$, the average
ellipticity of the population after a distortion $(g_1,g_2)$ is simply 
\begin{equation}
\left(\!\!
\begin{array}{c}\langle e_1\rangle\\\langle e_2\rangle\end{array}
\!\!\right)=
(1-\langle e^2\rangle)
\left(\!\!
\begin{array}{c}g_1\\g_2\end{array}
\!\!\right).
\label{eq:1mine2}
\end{equation}

\section{Shapelets}

The shapelets basis is described in Refregier
(\cite{Refregier2003}). It consists of the two-dimensional Cartesian
Gauss-Hermite functions, famous as the energy eigenstates of the 2-D
quantum harmonic oscillator:
\begin{equation}
B^{ab}(x,y)=k^{ab}\beta^{-1}
 e^{-[(x-x_c)^2+(y-y_c)^2]/2\beta^2} H^a(x/\beta) H^b(y/\beta).
\end{equation}
Here $(x,y)$ are coordinates on the image plane, $x_c$ and $y_c$ are
 the center of the expansion, $B^{ab}$ is the basis function of order
 $(a,b)$, $H^a$ is the Hermite polynomial of order $a$ and $\beta$ is
 a scale radius. $k^{ab}$ is a normalization constant chosen so that
 $\left(B^{ab}\right)^2$ integrates to one.

 Shapelets are a convenient basis set for describing astronomical
images because of the compact way in which various operators
(translation, magnification, rotation, shear) can be expressed as
matrices that act on the shapelet coefficients. Shapelets have a free
scale radius $\beta$ (the size of the Gaussian core of the functions),
and R03 shows how the coefficients transform under change of
$\beta$, and how to convolve objects with different scale radii.

To avoid introducing a preferred direction, the expansion should be
truncated in combined order $N=a+b$, not in $a$ or $b$ separately. (A
basis set truncated in $N$ is complete under rotation. Effectively,
such a truncation describes an image as a product of a circular Gaussian
with an inhomogeneous polynomial in $x$ and $y$ of order $N$. Rotation
of such an image will mix the $x^iy^j$ terms at constant $i+j$.)

The reason for choosing shapelets as a formalism for weak lensing
analysis is its ability to describe the main operations that a galaxy
image undergoes before it is registered at a telescope focal plane: in
reverse order, pixelation, convolution with a PSF, and distortion.

In this paper we concentrate on well-sampled (PSF FWHM at least 3--4
pixels), ground-based seeing-limited images. It remains to be seen to
what extent diffraction-limited PSFs, and undersampling, can be
handled with this formalism.

\subsection{Pixelation}

An image that is registered on a CCD is pixelated: the flux on the
surface of the detector is read out in binned form. Mathematically,
the flux is first boxcar smoothed (i.e., convolved with a pixel), and
then sampled at a spatial frequency of once per pixel. 
Therefore, if we fit a shapelet expansion to the binned image
$I(k,l)$ as a linear superposition of the basis functions $B^{ab}$:
\begin{equation}
I(k,l)=\sum_{a=0}^N\sum_{b=0}^{N-a} s_{ab} B^{ab}(k-x_c,l-y_c) 
\label{eq:pixellated}
\end{equation} 
then the shapelet coefficients $s_{ab}$ describe the pixel-convolved
image directly.  If a similar fit is made to a PSF image that is
pixelated in the same way, then the intrinsic, pre-seeing image is
exactly the deconvolution of the two shapelet expansions (apart from
the effects of noise, undersampling, and truncation of the shapelet
expansions).

\subsection{PSF convolution}

\begin{figure*}
\epsfxsize=\hsize\epsfbox{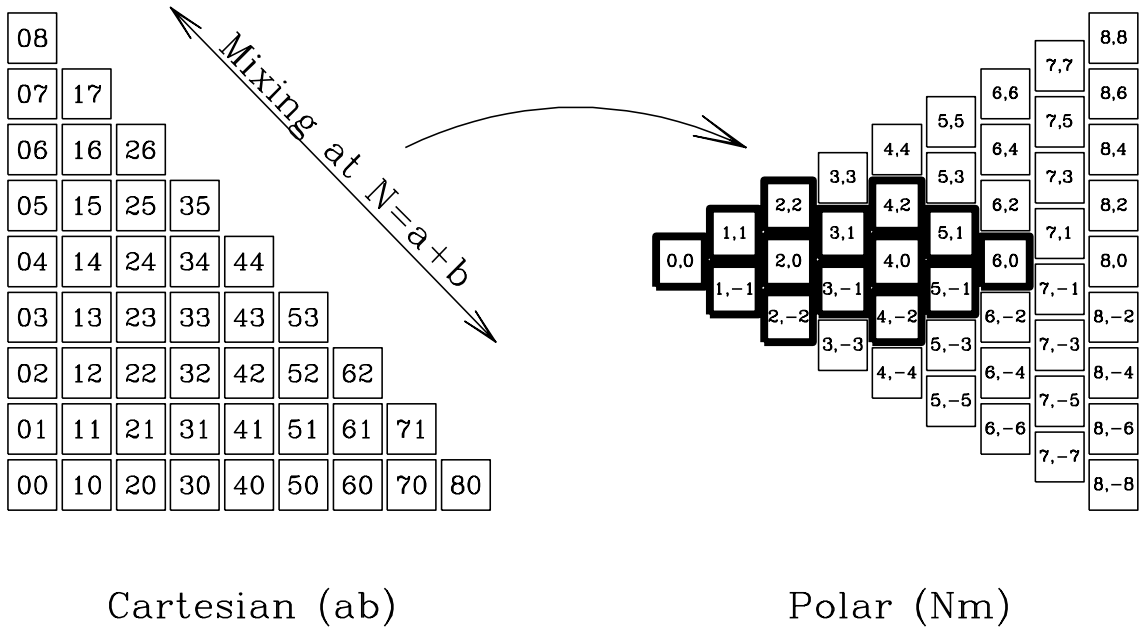}
\caption{The information used in the ellipticity determination, for
shapelet expansion to order $N=8$. On the left, the Cartesian shapelet
coefficients that are fitted to describe a source and its associated
PSF. On the right, the same information has been rearranged into a
polar shapelet expansion (the two may be transformed into one another
by appropriate mixing of the terms at order $N\equiv a+b$). The heavy
line indicates the polar shapelets from which the ellipticity is
estimated by a fit to eq.~\ref{eq:fittedmod}. Note the safety margin
at order $N-1$ and $N$, and the fact that only azimuthal orders
between $-2$ and $+2$ are fitted. }
\label{fig:fittedcoefs}
\end{figure*}

Convolution with the PSF can be expressed as multiplication with a PSF
matrix ${\bf P}_{a_1b_1a_2b_2}(\beta_\mathrm{in},\beta_\mathrm{out})$: if
the shapelet coefficients of the PSF are $p_{ab}$, and those of a
model source are $m_{ab}$, then their convolution is
\begin{eqnarray}
\nonumber
&\left(\sum_{a,b} p_{ab}B^{ab}_{\beta_\mathrm{psf}}\right)
\otimes
\left(\sum_{a,b} m_{ab}B^{ab}_{\beta_\mathrm{in}}\right)
\qquad\qquad\qquad
\\
&=
\sum_{a_1,b_1}\left(
\sum_{a_2,b_2} 
{\bf P}_{a_1b_1a_2b_2}(\beta_\mathrm{in},\beta_\mathrm{out})m_{a_2b_2}
\right)
B^{a_1b_1}_{\beta_\mathrm{out}} .
\end{eqnarray}
The subscripts on $B^{ab}$ identify the scale radius $\beta$, which
can be different for the three shapelet series involved: those for the
PSF, input model source and output result of the convolution. ${\bf
P}(\beta_\mathrm{in},\beta_\mathrm{out})$ convolves a shapelet expansion
with scale radius $\beta_\mathrm{in}$ with the PSF, resulting in a shapelet
expansion with scale radius $\beta_\mathrm{out}$. The coefficients of the PSF
matrix are
\begin{equation}
{\bf P}_{a_1b_1a_2b_2}(\beta_\mathrm{in},\beta_\mathrm{out})=
\sum_{a_3,b_3}
C_{a_1a_2a_3}^{\beta_\mathrm{out}\beta_\mathrm{in}\beta_\mathrm{psf}}
C_{b_1b_2b_3}^{\beta_\mathrm{out}\beta_\mathrm{in}\beta_\mathrm{psf}} 
p_{a_3b_3}
\label{eq:cnml}
\end{equation}
Here the elemental convolution matrix
$C_{nml}^{\beta_3\beta_1\beta_2}$ expresses the convolution of two
one-dimensional shapelets of scales $\beta_1$ and $\beta_2$ as a new
shapelet series with scale $\beta_3$. A recurrence relation for
$C_{nml}$ is given in R03.

\subsection{Ellipticity from shapelets}
\label{sec:ellshap}

Given a PSF and a PSF-convolved source, both expressed as shapelet
series, we determine the ellipticity of the source by modeling it as a
PSF-convolved, distorted circular source of arbitrary radial
brightness profile. This approach is similar to the one described in
Kuijken (\cite{Kuijken1999}), but it is more effective when expressed
in terms of shapelets.

A circular source of arbitrary radial brightness profile can be
written as a series of circular shapelets $C^n$ of the form
$c_0C^0+c_2C^2+c_4C^4+\ldots$, where the $c_n$ are free
coefficients. The $C^n$ (see Appendix) are normalized to have unit
integral over $x,y$, so $c_n$ gives the total counts in each
component. After distortion of such a source, it becomes an elliptical
source which, to leading order in ellipticity $e$, can be written as
\begin{equation}
(1+e_1{\bf S_1}+e_2{\bf S_2})(c_0C^0+c_2C^2+c_4C^4+\ldots) 
\label{eq:ellmodnopsf}
\end{equation}
where ${\bf S_i}$ are the first-order shear operators (see R03).  

This machinery allows us to write the model for the observed source as
\begin{equation}
{\bf P}\cdot(1+e_1{\bf S_1}+e_2{\bf S_2})(c_0C^0+c_2C^2+c_4C^4+\ldots),
\label{eq:ellmod}
\end{equation}
expressed as a set of shapelet coefficients that depend on the $e_i$
and the $c_n$.  Fitting this model to the shapelet coefficients of the
observed source (with their errors) yields the best-fit ellipticity
$(e_1,e_2)$ and the associated errors.

To improve the accuracy, we make two modifications: we add centroid
error parameters, and we only fit the model to a subset of the
shapelet expansions.  The centroid error parameters are included to
allow for a mismatch between the center of the object and its shapelet
expansion. If the PSF and/or the galaxy have some lopsidedness to
them, the center of their best-fit shapelet expansion may not be at
the flux-weighted center of the source (since our centering technique
simply requires the 01 and 10 components to be exactly zero). Hence
the centroid may move under convolution, which would spoil the fit. To
guard against this, instead of fitting Eq.~\ref{eq:ellmod} we fit a
model of the form
\begin{equation}
{\bf P}\cdot(1+e_1{\bf S_1}+e_2{\bf S_2}+d_1{\bf
  T_1}+d_2{\bf T_2})(c_0C^0\!+c_2C^2\!+\!\ldots\!+c_{N_c}C^{N_c})
\ \ \ 
\label{eq:fittedmod}
\end{equation} 
instead. The free translation terms $d_i{\bf T_i}$ in the model are
expressed in terms of the shapelet operators ${\bf T_i}$. 

The second modification is made to contain truncation errors. While
the shapelet basis is complete, and hence can describe any source
given enough terms, in practice the fact that the source is only
sampled in a finite set of pixels means that the expansion needs to be
truncated. Hence, except in very special cases, the PSF and galaxy are
not described perfectly by a truncated shapelet series. The missing
information propagates through the analysis, and is a source of
systematic error in the PSF convolution (as some PSF terms may be
missing) and in the calculation of the action of shear (since
shearing high-order shapelets generates also lower-order terms).

Truncation effects can be seen most clearly if we re-express the
shapelets in polar $(r,\theta)$ coordinates (they become Gaussians
times Laguerre polynomials of $r$---see BJ02, R03, Massey \& Refregier
\cite{Massey2005}). Polar shapelets are combinations of Cartesian
shapelets $B^{ab}$ of the same order $N=a+b$, whose angular dependence
is a pure sine or cosine of $m\theta$, for angular order $m$. The
order of the Laguerre polynomial is $N$, with $N\ge m$ and
$N+m$ even.

The translation and shear operators, when applied to a polar shapelet
of order $(N,m)$, generate terms at order $(N\pm1,m\pm1)$ and
$(N\pm2,m\pm2)$, respectively. If we truncate the shapelet series of
the best-fit circular model for the pre-seeing, pre-shear galaxy at
order $N_c$, then to be consistent the $m=1$ and $m=2$ series must be
truncated at order $N_c-1$ and $N_c-2$, respectively.

Note that we are never completely safe from truncation effects: in
particular, complex PSFs can in principle mix coefficients of all
orders. The problem is minimized, but not completely eliminated, by
adopting suitable scale radii so that the amount of information
carried in the high-order coefficients is small. We further include a `safety
margin' by setting $N_c=N-2$, in case the highest-order shapelet
coefficients are affected by PSF structure at even higher order. 
The scheme is illustrated for the typical case of $N=8$ in
Fig.~\ref{fig:fittedcoefs}. 
The highest-order polar shapelet coefficients that should be included
in the fit are $(N_c,0)$, $(N_c-1,\pm1)$ and $(N_c-2,\pm2)$.

\begin{figure}
\epsfxsize=\hsize\epsfbox{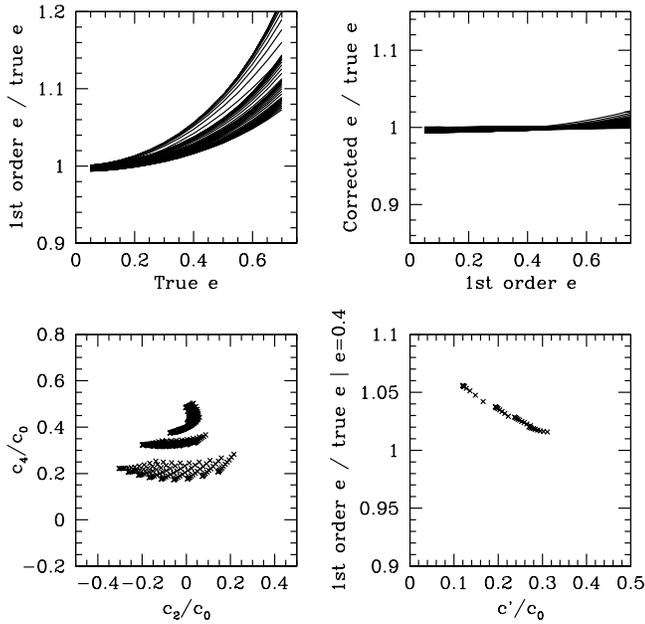}
\caption{Empirical calibration of the post-linear correction to the
measurement of $e$. Top left: fractional error on derived 1st-order
ellipticity $e$ for a range of different scale parameters and Sersic
indices. Top right: fractional error of the corrected
ellipticity. Bottom left: the coverage of the $c_2/c_0$, $c_4/c_0$
plane by the models. The four groups of points correspond, from top to
bottom, to Sersic index 4, 3, 2 and 1. The horizontal spread is mostly
a consequence of using different scale radii $\beta$. Bottom Right:
residuals on 1st-order ellipticity vs.\ the linear combination of
$c'/c_0=(0.085c_2+0.63c_4)/c_0$ for input ellipticity 0.4, showing
that this parameter drives the scatter. }
\label{fig:calshear}
\end{figure}

In summary, from the shapelet series for each source, up to Cartesian
order $N_\mathrm{max}$, we form a truncated polar shapelet series
including terms of order $(m=0, N=0,2,\ldots ,N_\mathrm{max}-2)$,
$(m=1,N=1,3,\ldots ,N_\mathrm{max}-3)$ and $(m=2,N=2,4,\ldots,
N_\mathrm{max}-4)$. The effect of the shear, translation and PSF
operators on circular basis functions up to order $N_\mathrm{max}-2$
is then calculated up to order $N_\mathrm{max}$, and the result
likewise converted to polar shapelets up to order $N_\mathrm{max}-2$.

Least-squares fitting the model to each source yields an ellipticity
estimate $(e_1,e_2)$, expressed as the shear that needs to be applied
to a circular source to fit the object optimally.

Performing the least-squares fit is straightforward to do
numerically. $\chi^2$ is a fourth-order polynomial in the fit
parameters $\{c_0,c_2,\ldots,c_{N-2},e_1,e_2,d_1,d_2\}$, and the
minimum can typically be found in a few Levenberg-Marquardt iterations
(Press et al.~\cite{Press1986}).

The errors on the shapelet coefficients for each source can be derived
from the photon noise, and these can be propagated through in the
$\chi^2$ function. The second partial derivatives of $\chi^2$ at the
best fit give the inverse covariance matrix, which can be inverted to
show the variance and covariances between the fit parameters.  This
results in proper error estimates on all parameters, in particular on
$e_1$ and $e_2$. In practice the errors on $e_i$ are only weakly
correlated with those on $d_i$ and $c_n$.

\section{Tests}

\label{sec:tests}
We now describe some elementary tests of this approach. 

\subsection{Test of ellipticity estimates}

For small ellipticity $e$ the linear shear operator provides a good
description of the action of the shear, but for larger $e$
higher-order corrections come into play, and these corrections depend
on the radial brightness profile. We have calibrated these corrections
empirically using a set of model sources that follow a Sersic
(\cite{Sersic1968}) distribution of brightness, of index 1--4. Each
source was sheared by varying amounts and encoded into shapelets using
a range of different scale radii. The first-order ellipticity estimate
$e_\mathrm{1st}$ derived by fitting a model of
eq.~\ref{eq:ellmodnopsf} was then compared to the true ellipticity
(see Fig.~\ref{fig:calshear}).  For small $e$ the correct ellipticity
is recovered, but at larger $e$ the discrepancy grows.  Fitting the
residuals versus the radial profile shape parameters $c_0$, $c_2$,
$c_4$, it turns out that the true ellipticity $e_\mathrm{true}$ can be
derived from the fitted 1st-order estimate by applying the correction
factor
\begin{equation} 
{e_\mathrm{true}\over e_\mathrm{1st}}
\simeq 1-e_\mathrm{1st}^2\left(-0.41+{0.085c_2+0.63 c_4\over c_0}\right).
\label{eq:corfac}
\end{equation}
The formula is valid to better than 1\%\ accuracy for $e<0.7$,
corresponding to axis ratios of nearly 6:1. 

Below we will, in fact, NOT apply this correction, because the errors
on $c_0$, $c_2$ and $c_4$ are typically so high, and (in the case of
small, barely resolved sources) correlated, that the correction factor
cannot be determined accurately.  Fortunately most galaxies in the sky
have ellipticity below 0.3, where the correction is below 3\%. If
necessary, the accuracy could be further increased by evaluating the
effect of shear on the shapelet basis functions to higher order in
eq.~\ref{eq:fittedmod}.

\subsection{Choice of Scale Radius}
\label{sec:scalerad}

A truncated shapelet expansion can only describe deviations from a
Gaussian over a particular range of spatial scales (which widens with
order $N$). For ellipticity determinations the outer parts of galaxy
and PSF images are most important (e.g., the classic second moments
depend on the 3rd moment of the radial profile), so there is some
advantage to taking as large a scale radius as possible.  On the other
hand, this radius should not be so large that the inner structure of
the source cannot be resolved.

\begin{figure}
\centerline{\epsfxsize=0.8\hsize\epsfbox{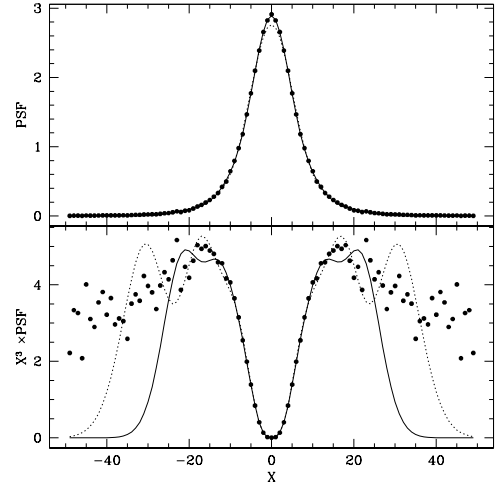}}
\caption{Different shapelet fits to a Moffat PSF $(1+\alpha r^2)^{-2}$
  with FWHM of 12 pixels. Top panel: the solid line shows a
  cross-section of an 8th-order fit, using the dispersion of the
  best-fit Gaussian as scale radius. The dotted line shows the
  corresponding fit for a scale radius that is a factor of 1.3 times
  larger. The lower panel shows the PSF multiplied by radius$^3$, in
  order to accentuate the residuals in the wings. Dots are the actual
  (monte-carlo realized) pixelated PSFs used in the simulations of
  \S\ref{sec:tests}. 
  Note how the increased scale radius makes for a much
  better fit in the outer regions, without a serious degradation in
  the core of the PSF.}
\label{fig:psffit}
\end{figure}

We have found that, for shapelet expansions up to order $N=8$ or
higher, taking a scale radius which is 1.3 times the dispersion of the
best-fit Gaussian works well for a range of model
PSFs. Fig.~\ref{fig:psffit} shows an example for a Moffat PSF with
index 2: both the core and the wings can be fitted adequately with
this choice of $\beta$.

\begin{figure}
\centerline{\epsfxsize=0.8\hsize\epsfbox{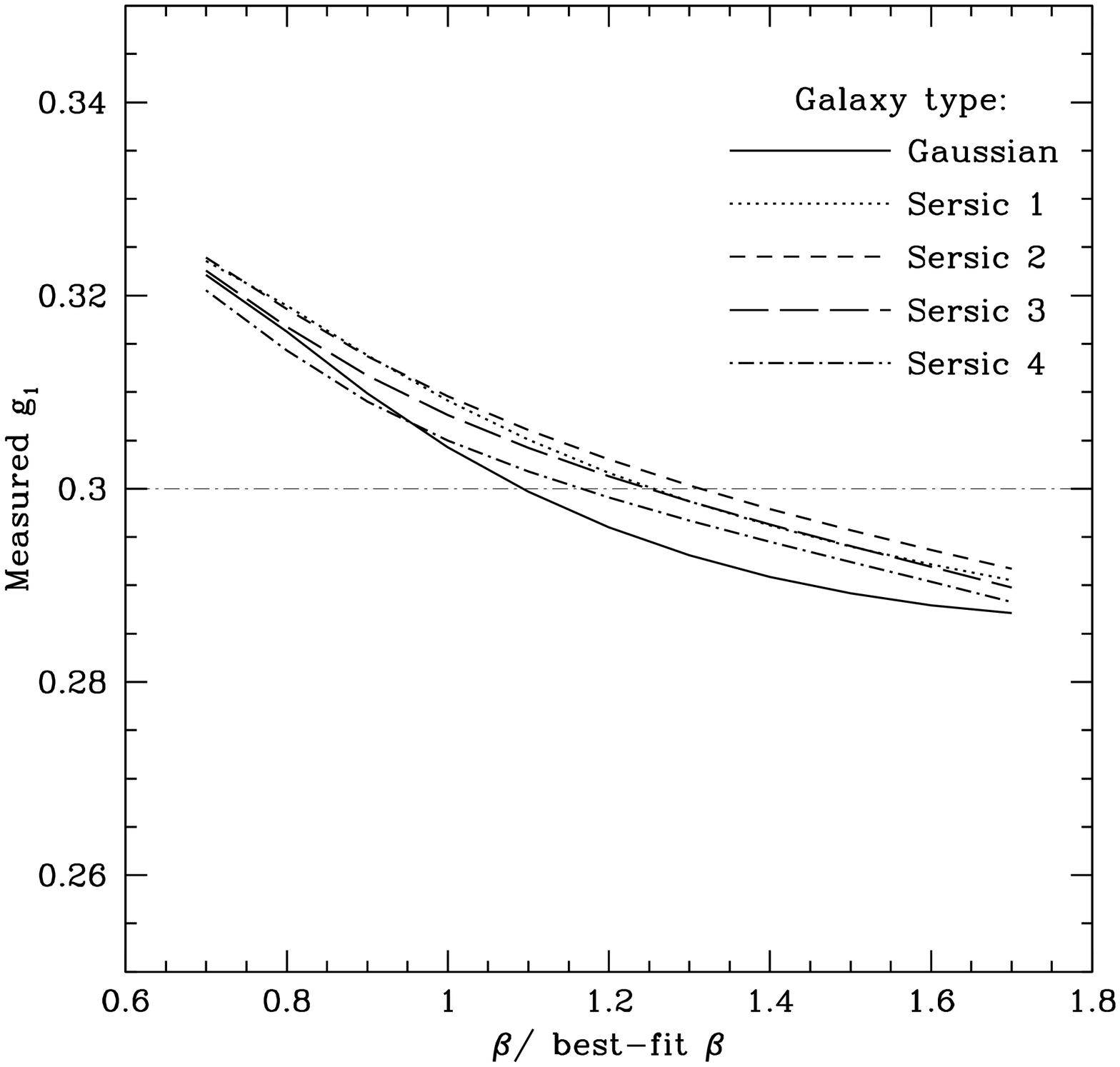}}
\caption{Illustration of the effect of the choice of scale radius on
the ellipticity measurement. Each curve shows, for a different galaxy
profile, how the derived $e$ depends on the choice of scale radius
(expressed as a multiple of the dispersion of the best-fitting round
Gaussian). Each model galaxy had an effective radius of 4 pixels and
ellipticity 0.3, and was convolved with a PSF of Moffat index 2 and
FWHM 8 pixels. The $\beta$ for convolved source and PSF are both
scaled by the same factor.}
\label{fig:varybeta}
\end{figure}

The impact of the choice of scale radius on ellipticity measurements
is shown in Fig.~\ref{fig:varybeta}, for typical images. The factor
1.3 represents a good compromise, though its exact value is
not critical.

\begin{figure*}
\centerline{\epsfxsize=0.9\hsize\epsfbox[18 144 592 532]{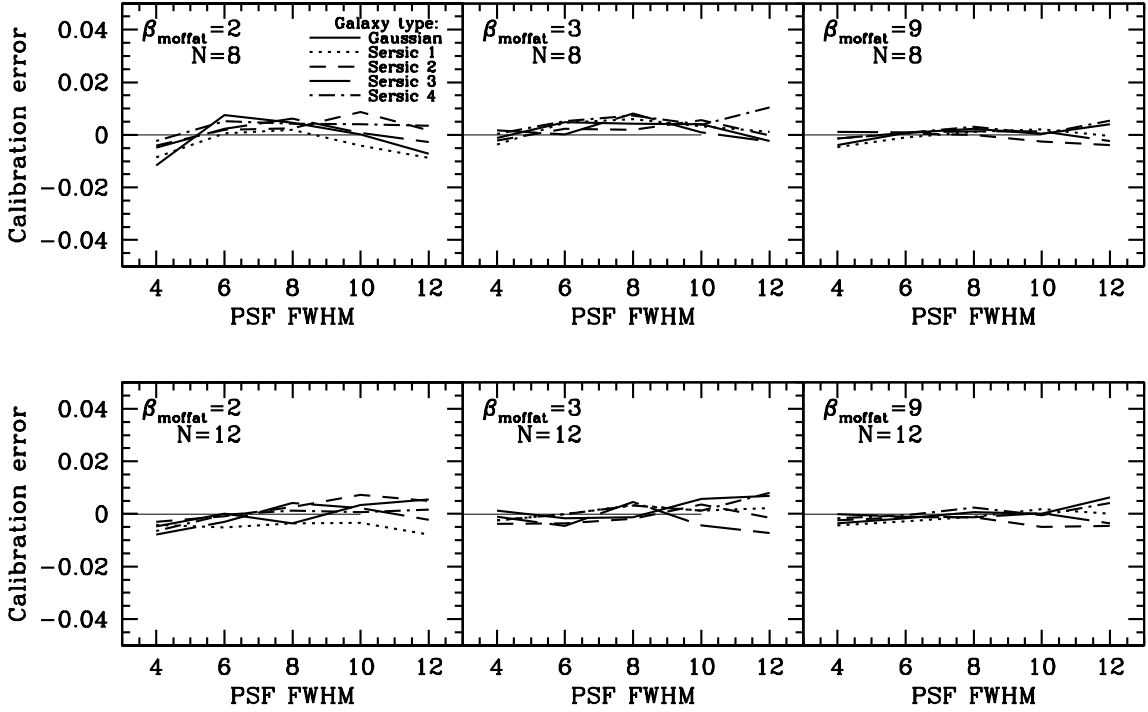}}
\caption{Fractional error in recovering a 10\% shear, using round
  PSFs. Each panel represents a different Moffat PSF; the rightmost
  panels are very nearly Gaussian. The simulated galaxies have
  effective radii of 4 pixels. Top row: 8th-order shapelets; bottom
  row: 12th-order shapelets.}
\label{fig:calfac}
\end{figure*}

\begin{figure*}
\centerline{\epsfxsize=0.9\hsize\epsfbox[18 144 592 532]{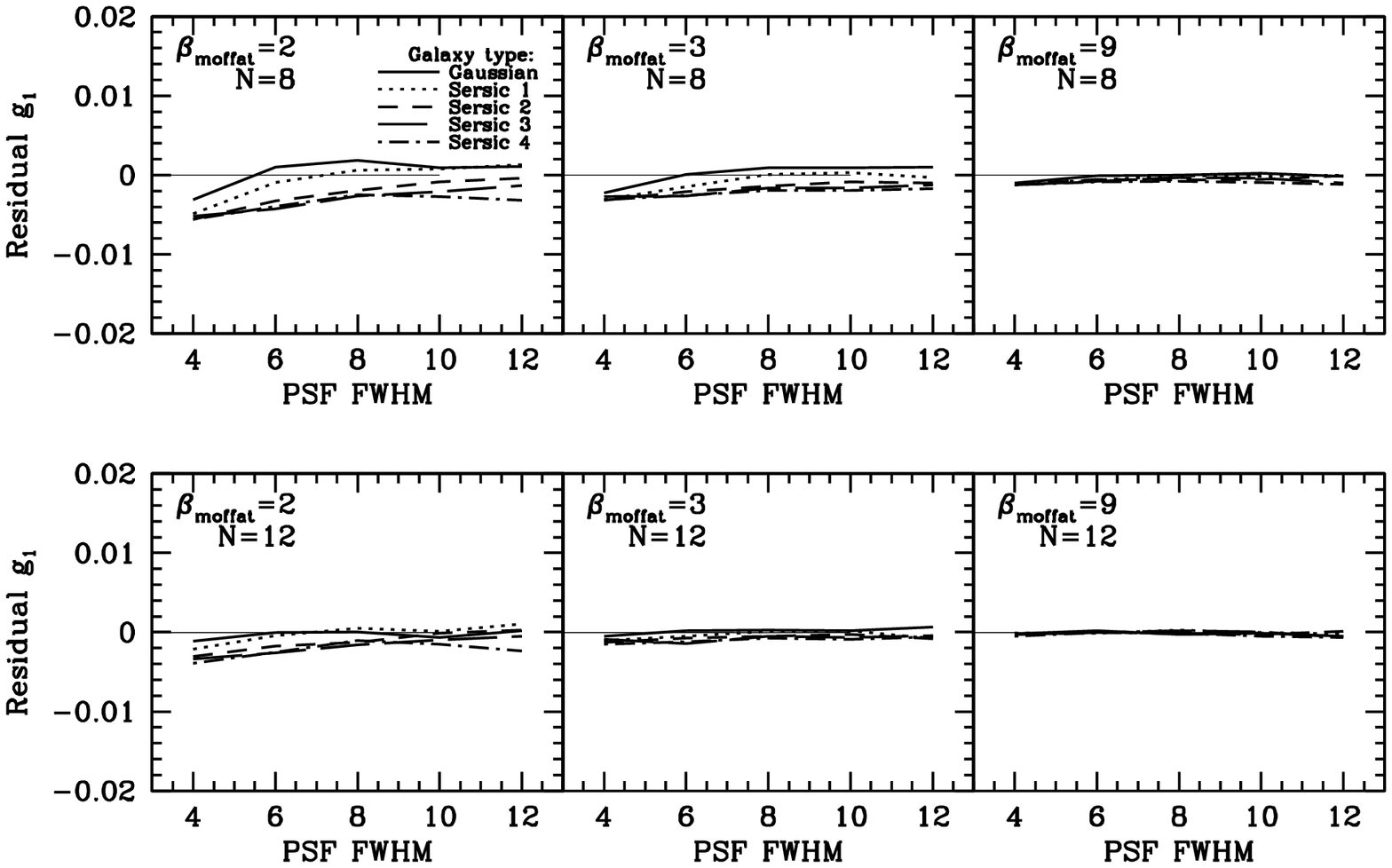}}
\caption{Residual shear after correction for an elliptical PSF (the
  same PSFs as fig.~\ref{fig:calfac}, sheared by 10\%).}
\label{fig:psfcor}
\centerline{\epsfxsize=0.9\hsize\epsfbox[18 144 592 532]{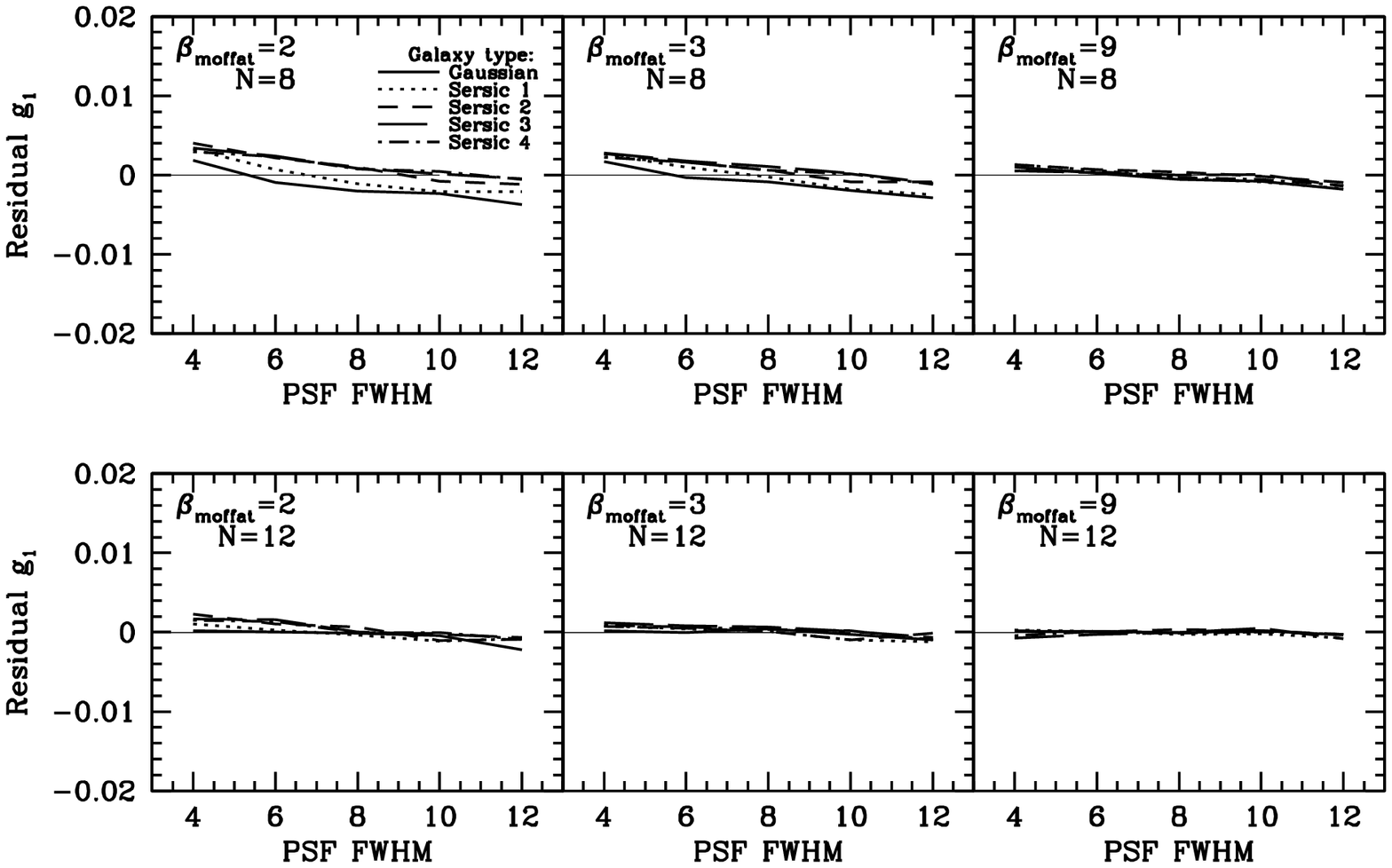}}
\caption{Residual shear after correction for a lopsided PSF (the
  same PSFs as fig.~\ref{fig:calfac}, but a third of their flux is
  displaced by (2/3)FWHM).}
\label{fig:asym}
\end{figure*}

\subsection{Test of PSF correction}
\label{sec:psftests}
Correcting for the effects of PSF convolution (dilution of ellipticity
by a round PSF, or biasing of ellipticity by an elongated one) is the
most critical part of weak lensing. The following tests show how well
the shapelets technique can do this. 

For the tests we generate simulated, high signal-to-noise (S/N)
sources. To be sure that pixelation effects are taken into account
properly, and that the PSF convolutions are done accurately, a
brute-force technique is used: each source is built of individual
`photons' that are drawn from a 2-D Sersic distribution, sheared if
required, then have a `PSF' displacement added to them, and finally
are added to the pixel in which they fall. We use 10 million photons
per source, which gives effectively noise-free images (S/N$\ga$1000).

We ran three sets of tests. In each case we explore galaxies with
Sersic laws $f(r)\sim\exp(-k r^{1/n_s})$ with indices $n_s=0.5$
(Gaussian), 1 (exponential), 2, 3 and 4 (de Vaucouleur), and use PSFs
with a Moffat profile $(1+\alpha r^2)^{-\beta_m}$ of index
$\beta_m=2$, 3 and 9 (nearly Gaussian). All galaxies are scaled to an
intrinsic effective radius of 4 pixels, and the PSF FWHM's range from 4 to 12
pixels. We use shapelet expansions to order $N=8$ and 12, and scale
radii of 1.3 times the dispersion of the best-fit Gaussian.

First, to test the `shear calibration' factor, we check how well we
can recover the shear of a galaxy that is sheared by $g_1=0.1$ and
convolved with a round PSF. Fig.~\ref{fig:calfac} shows the result:
any calibration error is at the sub-percent level; the worst results
are obtained for the most non-Gaussian PSFs ($n_s=4$). The noise in
the curves suggests that we are also limited by the accuracy of the
Monte-Carlo simulations of the galaxies, and by the numerics of the
software implementation of the method.

The second test shows to what extent PSF ellipticity pollutes the
shear estimate. The input PSFs were given an elipticity $e_1=0.1$
(axis ratio 0.82), and convolved with round galaxies. The recovered
$e$ is shown in Fig.~\ref{fig:psfcor}. The residual effect is at most
half a percent (worst case), which represents a correction of the PSF
ellipticity by a factor of better than 20. The best results are
obtained for lowish Sersic indices (below 2) and for PSFs with not too
extended wings (Moffat index 3 or higher). Perhaps surprisingly, the
larger the PSF (for the same galaxy size) the better the correction:
presumably this is a small sampling effect.

Finally, we introduced a lopsided PSF (by giving 1/4 of the photons an
extra offset of half the FWHM) and repeated the analysis. As can be
seen from Fig.~\ref{fig:asym}, this combination of dipole and
quadrupole PSF distortion can also be handled.

In all cases, taking the expansion to $N=12$ increases accuracy,
though not spectacularly. We conclude that the algorithm works:
shapelets provide a promising technique for measuring galaxy
ellipticities, and for correcting ellipticities for smearing by the
PSF.

\begin{figure*}
\epsfxsize=\hsize\epsfbox{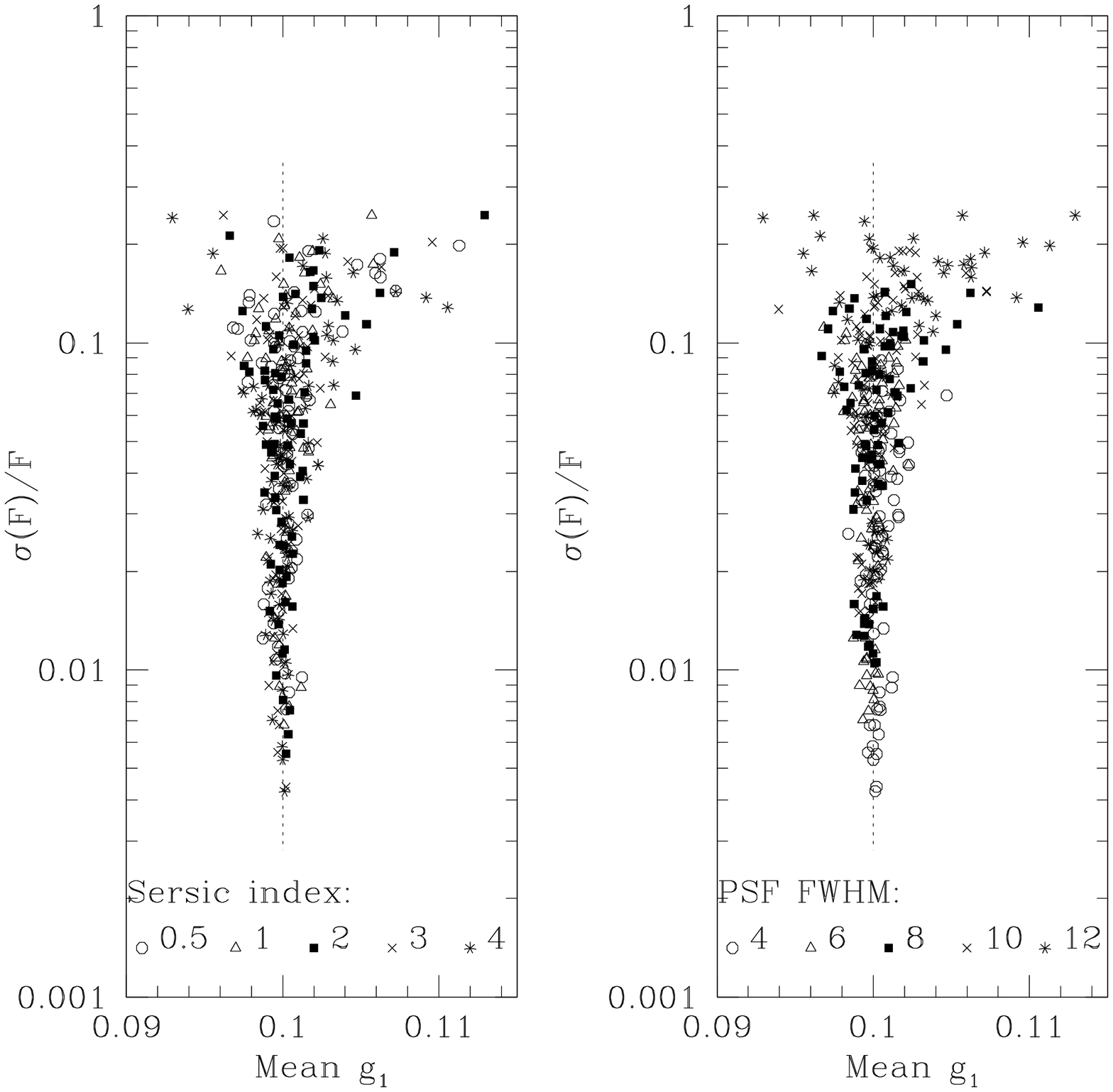}
\caption{The result of Monte Carlo simulations, in which many noise
  realizations of Sersic profile galaxies were run through the
  ellipticity-fitting procedure described in this Paper. Shapelet
  order $N=8$ was used throughout. Each plotted dot represents the
  average ellipticity of 2500 different noise realizations of the same
  galaxy image. The same data are plotted in both panels, but coded by
  different model parameters: the Sersic index on the left, and the
  PSF size on the right. The vertical axis shows the fractional
  scatter of the measured fluxes, $\sigma(F)/F$, of the sources, as
  determined by integrating their shapelet series. No trend of the mean
  shear with S/N is seen: noise leads to scatter but no bias.}
\label{fig:noisetests}
\end{figure*}

\begin{figure}
\epsfxsize=\hsize\epsfbox{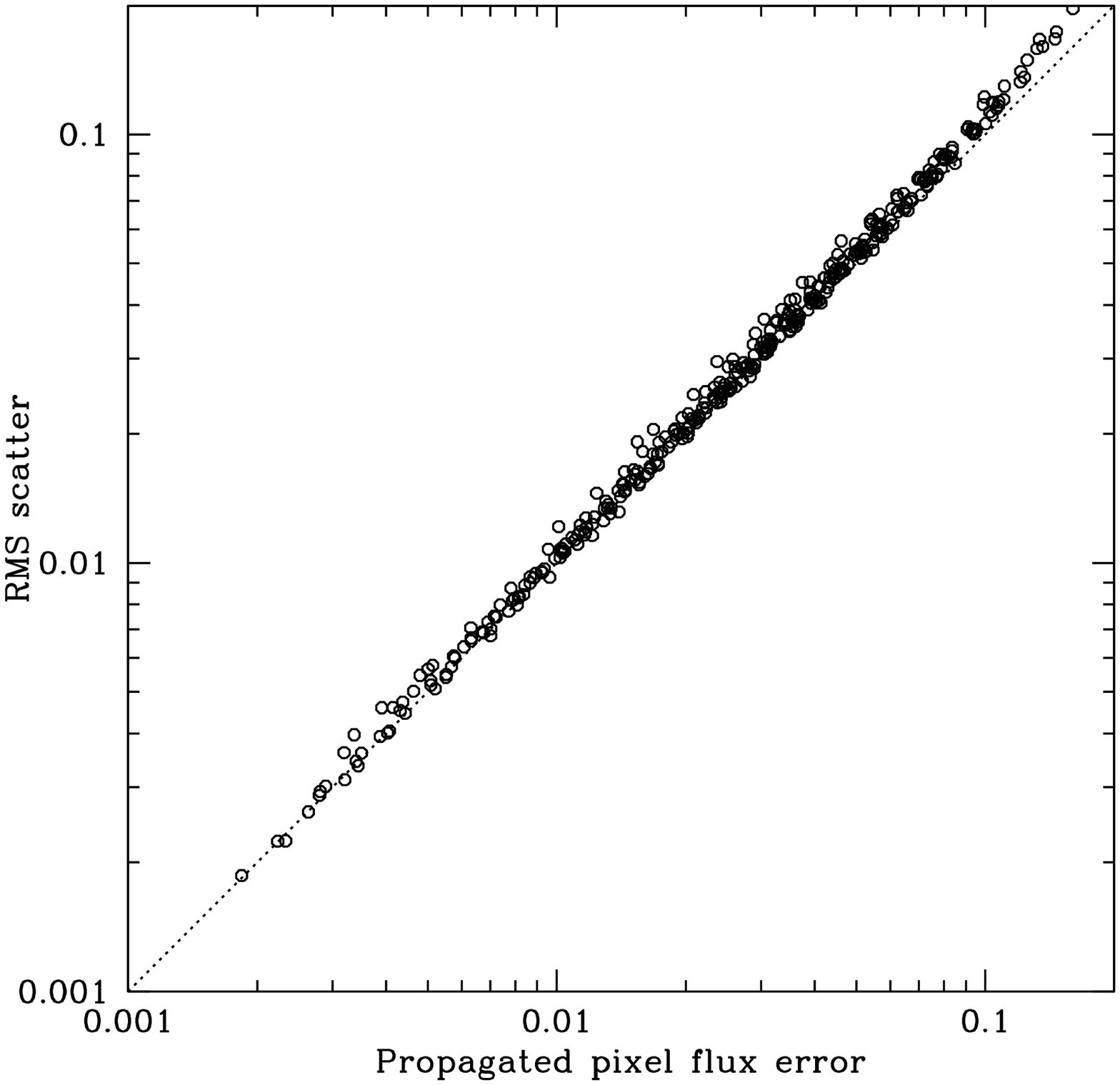}
\caption{Comparison between the scatter in ellipticity measurements
  from sets of 2500 random noise realizations, and the error predicted
  by propagating the pixel noise through the calculations.}
\label{fig:noiselevels}
\end{figure}

\subsection{Noise}

A final series of tests was used to check the behaviour of the
algorithm on noisy images. We added Poisson noise to simulations such
as those just described, and compared the result of many
realizations. The PSF FWHM ranged from 4 to 12 pixels, the galaxies'
effective radii were set to 4 pixels. The same ranges of Sersic
(0.5--4) and Moffat (2, 3, and 9) indices were used as above, and the input
shear was 0.1. Different levels of noise were added, roughly to span
the S/N range between 10 and 100.

The results are summarized in Fig~\ref{fig:noisetests}
and~\ref{fig:noiselevels}, which show that the noise causes a scatter
on the ellipticity estimates, but that this does not lead to a bias;
and that the propagated error estimates on the ellipticities are a
good measure for the rms scatter among the different realizations.

\section{The pipeline}

We have implemented the above ideas into a `shear pipeline'. It starts
from a fully reduced image, detects sources, determines the PSF and
its variation across the image, decomposes detected sources into
shapelets, and obtains a shear estimate for each object. The pipeline
consists of a number of stand-alone programmes that run in
sequence. The modules operate on source catalogues and FITS images,
and generate new catalogues and diagnostic plots. The pipeline can run
fully automatically.

\subsection{Detecting sources}

The first step in the reduction process is the detection of
sources. For this the SExtractor software (Bertin \& Arnouts
\cite{Bertin1996}) is used. A few basic parameters are measured during
extraction: position, the flux and its error, the FWHM,\footnote{FWHM
is determined by doubling the FLUX\_RADIUS parameter of SExtractor,
and not the FWHM\_IMAGE parameter which is prone to failure at low
fluxes.}  major and minor axis length and position angle, and source
quality flags. SExtractor is fast and effective, particularly for the
relatively high S/N sources that are used here.

\subsection{PSF maps from shapelets}

The PSF is determined from the stars in the image itself. We assume
that the pixel values are linearly related to intensity.

First, the stellar locus is determined from the plot of magnitude vs.\
FWHM in the standard way (e.g., KSB). We fit a circular Gaussian to
each star, and adopt the median dispersion from these fits, multiplied
by 1.3 (see \S\ref{sec:scalerad}) as the scale radius
$\beta_\mathrm{psf}$.

For all objects near the stellar locus a shapelet expansion is fitted
using $\beta_\mathrm{psf}$ as scale radius. The SExtractor centroid is
chosen as initial centroid for the expansion, but after the expansion
is completed the centroid is adjusted (using the 1st-order translation
operators of R03) until the coefficients of $B^{01}$ and $B^{10}$ are
exactly zero. Finding the required offsets involves solving a simple
linear equation. Each shapelet is finally normalized to unit integral,
by analytically integrating the counts in each shapelet term and
dividing by the total.

Once this is completed, we have a shapelet description of (candidate)
PSF objects scattered over the image. A map of the PSF variation
across the image can now be made by interpolating the shapelets. We
have found that a straightforward polynomial fit, coefficient by
coefficient, works well, though complex PSF variations may require
more sophisticated schemes such as weighted nearest neighbour averages
(Christen 2006), Pad\'e interpolants (Hoekstra \cite{Hoekstra2004}),
or even physically-motivated model fits (Jarvis \& Jain
\cite{Jarvis2004}; Jain, Jarvis \& Bernstein \cite{Jain2005}). During
the fitting of the variation of each coefficient over the image,
deviant points can be rejected, leaving a cleaned sample of PSF
objects.

The result of this step is a recipe for the shapelet coefficients of
the PSF at any point in the image.

\subsection{Shapelet encoding}

Once the PSF is determined, all other detected sources are expressed
as shapelets as well. As for the PSF objects, a shapelet expansion
centered on the SExtractor coordinates, is fitted directly to the
observed pixel values. The statistical errors on the pixel values are
propagated through the least-squares fitting, leading to errors (and
if desired, covariances) on the shapelet coefficients. In the case of
well-resolved shapelets and uniform noise level across the source, the
shapelet normalization is such that the rms error on each coefficient
is the same, and the correlation between errors on different
coefficients small.

Each source is encoded into shapelets with scale radius derived as
described in \S\ref{sec:scalerad}.  All pixel values within a radius
of $4\beta$ from the SExtractor centroid (at least 10 pixels) are used
in the fit.  For efficiency reasons in the shear estimation step, the
allowed $\beta$ values are quantized: allowed values are
$\beta=2^{n/8}\beta_\mathrm{psf}$, $n=0,1,2,\ldots$. After fitting,
the center of expansion for each object is shifted in the image plane
by means of the shapelet translation operators until the 01 and 10
coefficients are zero, as before.

This procedure describes the source as seen in the image plane, i.e.,
after it has been convolved with the PSF and pixelated. An alternative
approach also make sense: to convolve all basis functions with the
pixelated PSF, and fit the observed source image as a combination of
those. This yields a shapelet description for the intrinsic,
pre-seeing, object shapes (Massey \& Refregier \cite{Massey2005}). 
As long as the same procedure is followed for the sources and PSF, the
end result should be the same: the deconvolved image will be free of
pixelation and PSF.  We prefer our approach because it leaves the
covariance between the fitted shapelet coefficients small.

All sources are encoded to the same shapelet order as the PSF
(typically 8 or 12), in order to avoid signal-to-noise dependent
smoothing effects. For faint sources, the higher-order coefficients
will therefore be very noisy, but still unbiased.

\subsection{Shears from Shapelets}

With a description of the shape of each source and the corresponding
PSF, the next task is to determine the intrinsic shape parameters that
are needed for a weak lensing analysis.

As explained in Section~\ref{sec:ellshap}, we derive the ellipticity
as the shear that needs to be applied to a suitable round source in
order to fit the observed image. The fit is applied completely in
shapelet space. We use a scale radius equal to
$\left(\beta^2-\beta_\mathrm{psf}^2\right)^{1/2}$ for the intrinsic,
deconvolved, circular model galaxy, and normalize all sources to unit
flux before fitting so that the $c_n$ can be used as radial profile
shape parameters. Only sources with $\beta>\beta_\mathrm{psf}$ are
used. The $C_{nml}$ convolution coefficients need to be evaluated only
once per value of $\beta$.

\subsection{Cleaning the catalogue}

The resulting catalogue of sources with shape estimators needs to be
cleaned in order to remove sources which are affected by neighbours,
edge effects, poor fits, etc. We apply various cuts:

\begin{enumerate}
\item{\em SExtractor Flags\ }
We first exclude all objects for which SExtractor raised a flag (due
to neighbours, being close to the edge, saturation, etc.). 
\item{\em Unresolved and faint objects\ } Next size and S/N cuts are
applied to the catalogue: typically all objects with best-fit Gaussian
radius smaller than 1.1 times that of the PSF, and those with flux
less than 10 times the flux error (as measured by SExtractor
parameters FLUX\_AUTO and FLUXERR\_AUTO), are removed.
\item{\em Shape cuts\ }
For the next cut, for each source the fraction of power $F_n$ at each
order $n$ in the shapelet expansion is calculated. An 
unusually high amount of power at high order, particularly for odd
$n$, indicates a source whose shapelet expansion is affected by a 
neighbour. Thresholds are set for each $F_n$ based on the properties
of the ensemble population, above which sources are rejected.  Typical
values are $F_3>0.05$, $F_4>0.2$, $F_5>0.1$, $F_6>0.2$. By
construction $F_1=0$, but to filter out peculiarly lopsided sources a
maximum can be imposed on the distance by which the center of the
shapelet expansion had to be moved in order to set the 10 and 01
coefficients to zero. No cuts are applied to $F_2$
since that would be similar to a cut on image ellipticity.
\item{\em Radial profile cuts\ }
As a by-product of the shear estimation, radial profile parameters
$c_0\ldots c_{N-2}$ are generated. These can be used to further
excise peculiar objects. If the shapelet scales are chosen properly
and the shear fit worked well, most of the flux of the pre-seeing,
pre-shear model source should be contained in the $C^0$
term. Catastrophic failures of the fit, or problems with the setting
of the shapelet scale, can be identified as peculiar values of
$c_0$. Requiring $|c_0-1|<0.5$ filters out such cases.
\end{enumerate}

\subsection{Average shear}

Once individual estimates have been obtained for each source these
need to be combined in some way to generate a shear estimate.

Conceptually the simplest methods are (i) to average the $e_i$ and
divide by $(1-\langle e^2\rangle)$ (eq.~\ref{eq:1mine2}), and (ii) to
identify the mode of the ellipticity distribution (provided it is
centrally peaked), which identifies the intrinsically round galaxies.

A better technique is to form a weighted mean, where the weight is
driven by the amount of information about the shear field each source
provides. The scatter in the ellipticity measurements of sources is
due to the intrinsic dispersion in shapes $s_e$, and to measurement
errors. The latter can be estimated by propagating the noise in each
image through the fitting procedure, and the former can be estimated
as the excess variance of $e$ in the source population. We therefore
adopt individual weights $w=(s_e^2+\sigma_1^2+\sigma_2^2)^{-1}$ when
forming the mean of all measured ellipticities. 

The same weighting is then applied to determine the value of
$(1-\langle e^2\rangle)$ in eq.~\ref{eq:1mine2}. So the shear estimate
is determined as
\begin{equation}
w=(s_e^2+\sigma_1^2+\sigma_2^2)^{-1} 
\label{eq:wts}
\end{equation}
\begin{equation}
\langle e^2\rangle=\displaystyle{\overline{w(e_1^2+e_2^2)} -
  \overline{w(\sigma_1^2+\sigma_2^2)} \over \overline{w}}
- \left({\overline{we_1}\over\overline{w}}\right)^2
- \left({\overline{we_2}\over\overline{w}}\right)^2
\label{eq:e2}
\end{equation}
\begin{equation}
g_i=\displaystyle{1\over 1-\langle e^2\rangle}{\overline{we_i}\over \overline{w}}
\end{equation}
The value of $s_e^2$ in eq.~\ref{eq:wts} can be iteratively adjusted
to equal $\langle e^2\rangle$ from eq.~\ref{eq:e2}, though its precise
value is of little consequence.

Depending on the form of the intrinsic shape distribution of galaxies,
different weightings are optimal: for example, for a very peaked
distribution of ellipticities higher weight can be given to nearly
round sources, whereas for a top-hat distribution the sources with
large ellipticity carry more information---for a discussion see
BJ02. A problem with weight factors that depend on $e$ is that the
centroid needs to be found first as it is the intrinsic, pre-shear
ellipticity that counts. The weight adopted above is optimal for a
Gaussian distribution of intrinsic ellipticities.

In the tests below we will compare the weighting scheme just
described, and a simple unweighted median. To the extent that the
median identifies the center of the distribution of ellipticities of
the source population, i.e., the intrinsically round sources, no
$(1-\langle e^2\rangle)$ correction needs to be applied to the median.

\section{Tests on STEP1 data}
\label{sec:step1}

\begin{figure}
\centerline{\epsfxsize=0.9\hsize\epsfbox[18 144 300 718]{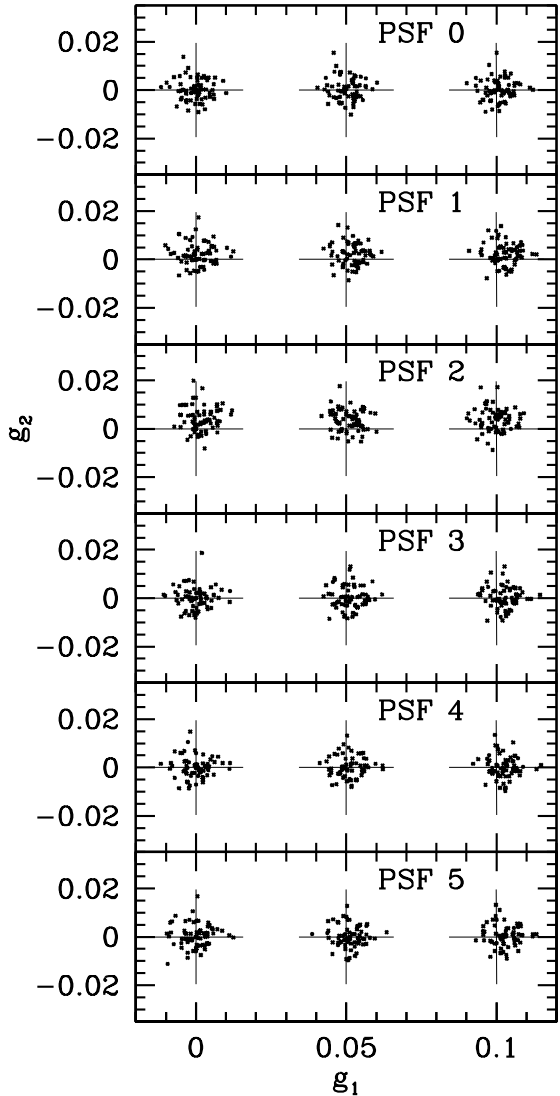}}
\caption{Results from the STEP1 simulations, which are based on
  modeling of the optics of the CFHT. Each plotted point represents
  the average ellipticity of about 2200 sources in one STEP1 image.
  Shapelets to order $N_\mathrm{max}=8$ were used. PSFs 0 to 5 are,
  respectively, round, with coma, with astigmatism, with defocus, and
  with 3rd and 4th-order astigmatism, and results for images with
  applied shears of 0, 0.05 and 0.1 are shown as three clusters of
  points in each panel. The correction factor $1-\langle e^2\rangle$
  is about 0.93 in all cases. Results from cluster to cluster are not
  statistically independent, but within each cluster of points they
  are. }
\label{fig:step1}
\end{figure}

\begin{figure}
\centerline{\epsfxsize=0.9\hsize\epsfbox[18 144 300 718]{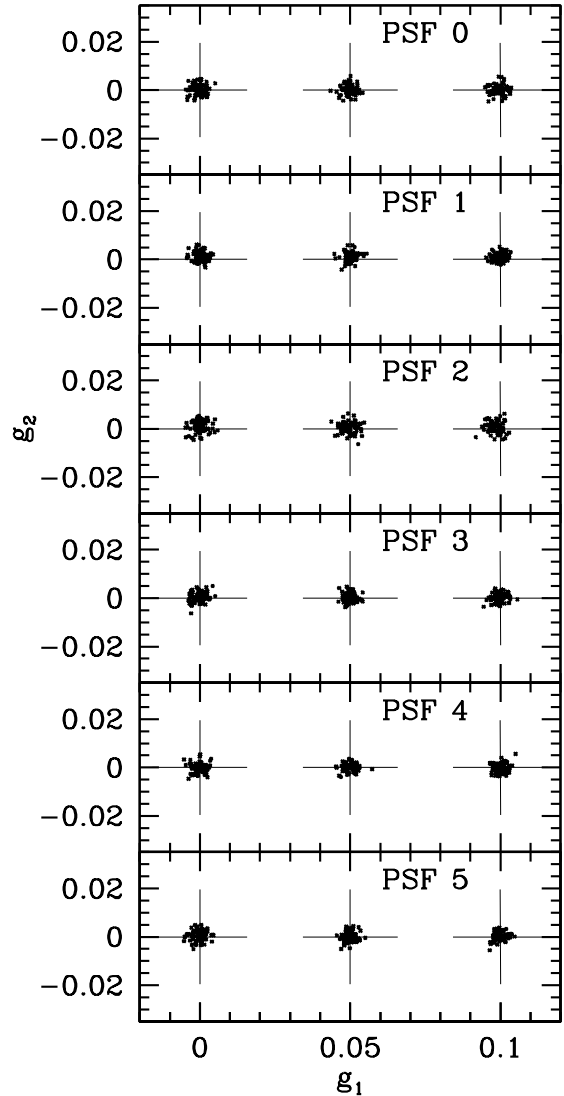}}
\caption{As Fig.~\ref{fig:step1}, but now the median ellipticity is
  used as shear estimator. The STEP1 galaxy ellipticity distribution
  is very peaked, which makes the median a very efficient estimator of
  the center of the distribution.}
\end{figure}

A rather realistic test of the whole procedure is provided by the
`Shear Testing Programme' (STEP, Heymans et
al.\ \cite{Heymans2005}). Phase 1 of STEP has produced a large set of
realistic simulated images across which a constant shear (in $g_1$)
and PSF smearing has been applied. The PSFs mimic realistic optical
aberrations. These images provide an important test, since unlike the
tests presented in \S\ref{sec:tests}, the pre-seeing sources do not
have a constant ellipticity with radius. Furthermore, the STEP images
include photon noise, and the sources may overlap. The PSFs are also
somewhat smaller (in pixels) than in the tests of \S\ref{sec:ellshap}:
scale radii used range from 2.2 to 3.2.

An analysis of the STEP1 data with an earlier version of this software
was reported in Heymans et al.\ (\cite{Heymans2005}). The main
improvement in the implementation since then has been the use of polar
shapelets in the shear determinations, which allows truncation effects
to be curtailed properly, and the use of larger scale radii as
discussed in \S\ref{sec:scalerad}.

Results of the use of the present pipeline on the STEP1 data are shown
in Fig.~\ref{fig:step1}. For each of three $g_1$ shear values, and six
different PSFs, 64 separate simulated images were run through the
pipeline, each image yielding shear estimates based on about 2200
galaxies. Table~\ref{tab:stepresults} summarizes the results per PSF
in terms of a multiplicative correction factor $m$, and an additive
offset $c$.  It can be seen that in general the method suffers from
very little bias. For the non-elliptical PSFs (0,3,4,5), the recovery
is perfect within the noise ($m=1$, $c=0$), which indicates that the
correction for dilution by PSF smearing works. On the other hand, for
the comatic (1) and elliptical PSF (2) there is a residual additive
term, of nearly 0.003 and 0.005 in shear respectively. This result is
consistent with that of the simulations in \S\ref{sec:psftests}. In
addition the elliptical PSF suffers from a multiplicative bias of
nearly 4\%. The origin of this discrepancy is not clear at the moment.

\begin{table}
\begin{tabular}{clrrrlrrr}
\multicolumn{1}{c}{}&&
\multicolumn{3}{c}{Weighted Average}&&
\multicolumn{3}{c}{Median}\\
\multicolumn{1}{c}{PSF}&&
\multicolumn{1}{c}{$m_1$}&
\multicolumn{1}{c}{$c_1$}&
\multicolumn{1}{c}{$c_2$}&&
\multicolumn{1}{c}{$m_1$}&
\multicolumn{1}{c}{$c_1$}&
\multicolumn{1}{c}{$c_2$}\\
0 && 0.995 & -0.02 & 0.03 && 0.982 & -0.03 & 0.01 \\
1 && 1.005 & 0.06 & 0.22 && 0.981 & 0.01 & 0.12 \\
2 && 0.963 & 0.19 & 0.38 && 0.967 & 0.02 & 0.07 \\
3 && 1.009 & 0.00 & 0.03 && 0.984 & -0.02 & 0.02 \\
4 && 1.010 & -0.01 & 0.02 && 0.986 & -0.02 & 0.01 \\
5 && 1.012 & -0.01 & 0.03 && 0.988 & -0.04 & 0.02 \\
\end{tabular}
\caption{Summary of the results of the STEP1 simulations. For each
  PSF, the slope $m_1$ and intercepts $c_i$ of the best-fit linear
  relation between input and recovered shear are shown. The $c_i$ have
  been multiplied by 100 for clarity. As no input
  $g_2$ distortion was applied in the STEP1 simulations, $m_2$ cannot
  be measured.}
\label{tab:stepresults}
\end{table}

\begin{figure}
\epsfxsize=\hsize\epsfbox{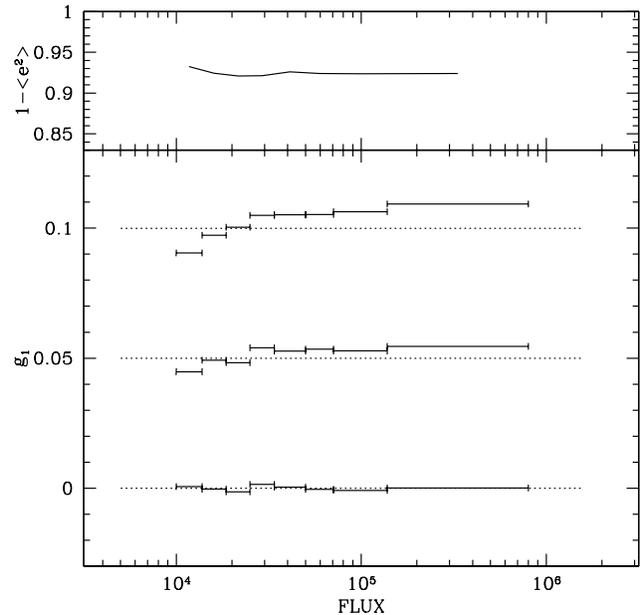}
\epsfxsize=\hsize\epsfbox{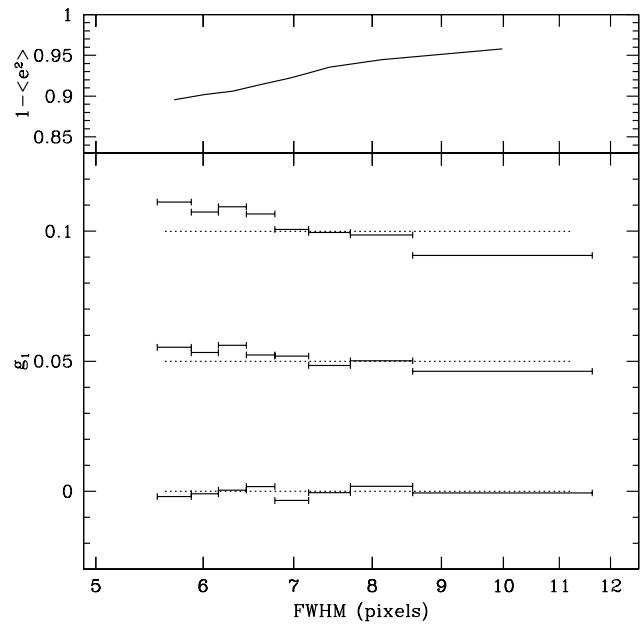}
\caption{The recovered shear for the STEP1 simulations for PSF0 and
  input shears of 0, 0.05 and 0.1, split up by source brightness (top)
  and size (bottom). In both cases there is a systematic, so far
  unexplained trend. The upper panel in each plot shows the
  ellipticity dispersion correction factor derived for and applied to
  each bin.}
\label{fig:magradtrends}
\end{figure}

The STEP1 analysis revealed that many methods show a small systematic
trend between the error in the derived shear and the magnitudes and
sizes of the objects (Heymans et al.~\cite{Heymans2005}). The
technique presented in this Paper is no different, as illustrated in
Fig.~\ref{fig:magradtrends}. These trends are still a puzzle, but it
is clearly important to trace their origin and further improve the
accuracy of the shear measurement.. Possible causes include
ellipticity-dependent incompleteness in the catalogues, problems
estimating the intrinsic ellipticity dispersion $\langle e^2\rangle$,
magnitude- or size-dependent neighbour contamination, or residual
systematic issues in the method itself. Further work addressing these
issues is on-going.

\section{Comparison with other techniques}

The method described here has several advantages. It goes beyond the
KSB description of PSF anisotropy as a convolution with a very compact
PSF, and is in principle applicable to all PSF shapes. The correction
for the PSF is performed in a single step, which avoids the need to
separate the effect of the PSF into an anisotropic part that shifts
the ellipticities, and a round part that causes a dilution of the
source ellipticities. The forward-fitting approach of a PSF-convolved
model for the intrinsic galaxy shapes to the observations allows error
propagation. The fact that the ellipticities derived are 'geometric'
in the sense of BJ02 (i.e., they represent the shear to apply to a
round object in order to fit the source) means that there is no need
to derive a higher-order 'shear polarizability', but instead the
response of an ensemble of sources to shear can be predicted simply
from the dispersion in ellipticities.

The use of shapelets in this method is not essential, but does help to
speed up the calculations, and gives a natural framework for isolating
the $m=2$ components that carry the ellipticity information about a
galaxy. It also allows the source lists to be filtered efficiently
based on objective shape criteria, and gives a robust way of
interpolating the PSF shape across an image. In cases where the PSF
cannot we well-described by a truncated shapelet expansion (for
example, poorly-sampled space-based observations) it is possible to
extend the technique, by performing the PSF convolution of
eq.~\ref{eq:fittedmod} in pixel space instead of in shapelet space
(Kuijken \cite{Kuijken1999}).

As was shown by the STEP1 project (Heymans et al.\ \cite{Heymans2005})
a variety of techniques can be used to derive shears from ground-based
image data, with residual errors around the 1\% level. While the focus
of STEP1 was on different variations of the KSB method, we have shown
here that the shapelets technique can do as well. As KSB is expected
to hit a fundamental level of systematic error (because of its good
but imperfect description of the PSF effects), which may well be
inadequate for the next generation of weak shear surveys, it is
worthwhile to look to higher-order methods.

The approach we have taken here differs in several ways from that of
Refregier \& Bacon (\cite{RefregierBacon2003}) and Massey \& Refregier
(\cite{Massey2005}), which is also shapelets-based. We perform the
shapelet expansions to a fixed order, and prefer not to introduce
signal-to-noise dependent thresholds that may lead to biases in the
derived shears (S/N dependent truncation and averaging do not
commute). Our non-iterative procedure is also much faster. Our
shapelet expansions describe the observed, post-seeing images, which
means that the coefficients are statistically almost independent
(whereas shapelet coefficients of the deconvolved shapes are
necessarily correlated), ensuring that they are virtually unaffected
by truncation of the series. We explicitly allow for errors in the
centroiding of the sources in our ellipticity estimates, and
forward-fit the observed images to account for PSF effects rather than
deconvolving them using a truncated PSF shapelet expansion. Finally,
rather than modeling the response of a (model or empirical) population
of images to a shear, we derive `geometric' ellipticities (BJ02) for
each individual source, and average these only at the end.

Compared to the Bernstein and Jarvis (BJ02) technique, the two main
differences are the way ellipticity is measured, and the way PSF
effects are handled. BJ02 derive geometric ellipticities by fitting an
elliptical Gaussian form to the observed images. This is similar to,
but not quite the same as, a low-order shapelet fit as used in this
paper. For sources with constant-ellipticity isophotes the BJ02 method
is exact, whereas ours is accurate only to $o(e^2)$ (see
Fig.~\ref{fig:calshear}). The benefit is a much faster numerical
convergence. BJ02 correct the ellipticities for PSF convolution by
means of a description of the PSF as a Gauss-Laguerre expansion
(equivalent to polar shapelets).  Optionally, the images can be
convolved with a rounding kernel before the ellipticities and PSF are
measured, in order to improve the accuracy of the PSF model. Thus the
main difference with the approach here is that BJ02 first derive a
post-seeing ellipticity, which is then corrected for the PSF; here we
forward-fit the intrinsic ellipticity in one step.
BJ02 separate the effects of PSF anisotropy (ellipticity
bias) and circularly-symmetric smearing (ellipticity dilution). They
correct for the latter by assuming an intrinsic light profile of the
source that is a perturbation about a Gaussian, expanded up to the
kurtosis.  The fact that here we model the full intrinsic radial
profile of the source as higher-order shapelets should provide higher
accuracy.

The Kaiser~(\cite{Kaiser2000}) technique is a more sophisticated
kernel convolution technique, in which a convolution kernel is
constructed which turns an image into one for which the effect of
pre-seeing shear on all sources is known exactly. This allows one to
find the shear that makes the source ellipticity distribution
isotropic---this is then the opposite of the amount by which the
population was sheared on its way to the telescope. The method is in
principle exact, and operates in pixel space. It appears to have
received relatively little use thus far (Wilson et al.\
\cite{Wilson2001a}, \cite{Wilson2001b}; Dahle et al.\
\cite{Dahle2002}).

\section{Conclusions}

Shapelets provide a neat framework in which to describe the
transformations that an astronomical source image undergoes until it
is registered on a detector. Gravitational shear, convolution with a
PSF, and pixelation can all be modeled within the shapelets
formalism.  All these elements can be combined into an efficient
algorithm for extracting image ellipticities that can be used for
accurate gravitational lensing shear measurements.

The implementation of these techniques into a working pipeline is
presented in this paper. Tests show that the pipeline is able to
recover input gravitational shears with very small calibration error
(of the order of a percent) and PSF residual (better than a factor of
30 in PSF ellipticity).

It remains to be seen to what extent this approach can be applied
successfully to diffraction-limited PSFs, which cannot be described
easily with a shapelet expansion. A different set of basis functions
for the expansion might be the answer. Further possible improvements
are also under investigation.

\begin{acknowledgements}

Ludo van Waerbeke provided the simulated `STEP' images used in
\S\ref{sec:step1}.  The Leids Kerkhoven Bosscha Fonds is thanked for
travel support.

\end{acknowledgements}

\appendix

\section{Circular Shapelets}
The Cartesian shapelets $S^{ab}$ at order $n=a+b$ can be written as
inhomogeneous polynomials in $x$ and $y$ of combined order $n$ times a
circular Gaussian. The leading-order term for $S^{ab}(x,y)$ is (R03)
\begin{equation}
2^nkx^ay^b e^{-r^2/(2\beta^2)}
\end{equation}
where $k$ is a constant that is independent of order, and $r^2=x^2+y^2$.
The circular shapelet at (even) order $n$ is the unique linear
combination of the $S^{ab}$ with $a+b=n$ 
that depends only on $r$: to leading order in $r$
\begin{eqnarray}
\nonumber
C^n(r)
&\equiv k'\sum_{i=0,2\ldots,n}
\left(\!\!
\begin{array}{c}n/2\\i/2\end{array}
\!\!\right)
S^{i,n-i}(x,y)\\
&= 2^nkk'(x^2+y^2)^{n/2} e^{-r^2/(2\beta^2)},
\label{eq:circshape}
\end{eqnarray} 
where $k'$ is chosen to normalize $C^n$ to unit integral over the $xy$
plane.  From the fact that rotation of Cartesian shapelets only mixes
terms of the same order $n=a+b$ it follows that the linear combination
in eq.~\ref{eq:circshape} is circularly symmetric at all orders of $r$.

\end{document}